\documentclass[pdflatex,sn-basic,iicol]{sn-jnl}

\usepackage{graphicx}%
\usepackage{multirow}%
\usepackage{amsmath,amssymb,amsfonts}%
\usepackage{amsthm}%
\usepackage{mathrsfs}%
\usepackage[title]{appendix}%
\usepackage{xcolor}%
\usepackage{textcomp}%
\usepackage{manyfoot}%
\usepackage{booktabs}%
\usepackage{algorithm}%
\usepackage{algorithmicx}%
\usepackage{algpseudocode}%
\usepackage{listings}%

\begin{document}

\title[Simulating Rotating Newtonian Universes]{Simulating Rotating Newtonian Universes}

\author*[1,2]{\fnm{Balázs} \sur{Pál}}\email{pal.balazs@ttk.elte.hu}

\author[3]{\fnm{Tze} \sur{Goh}}
\equalcont{These authors contributed equally to this work.}

\author[4]{\fnm{Gábor} \sur{Rácz}}
\equalcont{These authors contributed equally to this work.}

\author[3]{\fnm{István} \sur{Szapudi}}
\equalcont{These authors contributed equally to this work.}

\affil*[1]{\orgdiv{Department of Physics of Complex Systems}, \orgname{ELTE E\"{o}tv\"{o}s Loránd University}, \orgaddress{\country{Hungary}}}
\affil[2]{\orgdiv{Institute for Particle and Nuclear Physics}, \orgname{HUN-REN Wigner Research Centre for Physics}, \orgaddress{\country{Hungary}}}
\affil[3]{\orgdiv{Institute for Astronomy}, \orgname{University of Hawai'i at M{\=a}noa}, \orgaddress{\city{Honolulu}, \state{HI}, \country{USA}}}
\affil[4]{\orgdiv{Department of Physics}, \orgname{University of Helsinki}, \orgaddress{\country{Finland}}}

\abstract{We present the results of a novel type of numerical simulation that realizes a rotating Universe with a shear-free, rigid body rotation inspired by a G\"{o}del-like metric. We run cosmological simulations of \textbf{unperturbed glasses} with various degrees of rotation in the Einstein--de Sitter and the $\Lambda$CDM cosmologies. To achieve this, we use the \texttt{StePS} N-body code capable of simulating the infinite Universe, overcoming the technical obstacles of classical toroidal (periodic) topologies that would otherwise prevent us from running such simulations. Results show a clear anisotropy between the polar and equatorial expansion rates with more than $1~\%$ deviation from the isotropic case for maximal rotation without closed timeline curves within the horizon, $\omega_{0} \approx 10^{-3}$ Gyr$^{-1}$; a considerable effect in the era of precision cosmology.}


\maketitle

\section{Introduction} \label{sec:introduction}

Recent analyses of cosmological surveys indicate a statistically significant anisotropy in the expansion rate of the Universe (e.g., \cite{migkas2020,mohayaee2021,mcconville2023}), suggesting new aspects to cosmology beyond the Lambda Cold Dark Matter ($\Lambda$CDM), the current concordance model. One possible explanation for a direction-dependent anisotropy could be a cosmological rotation of the Universe itself.

The idea of a \textit{rotating Universe} is not new; for a review, see \cite{obukhov2000}. While numerical approaches of a rotating Universe were considered previously by e.g. \cite{buser2013}, it has never been studied using cosmological simulations in a full 3D setting. The concept of cosmic rotation is also not without merit: rotation is pervasive in the Universe, from microscopic scales to large astronomical structures. Planets, stars, galaxies, and even galaxy clusters all exhibit rotational motion.

Although \cite{lanczos1924} was the first to describe the Einstein Static Universe in a rotating (cylindrical) setting, \cite{gamow1946} was the first to propose the intriguing idea that the rotation of galaxies might be induced by an intrinsic, global rotation of the Universe itself. The first groundbreaking work to lay out a cosmological model of a rotating Universe was made by \cite{godel1949}, introducing a homogeneous, swirling dust solution to the Einstein's Field Equations (EFE) with a negative cosmological constant.

\cite{godel1949} proposed this model as a pedagogical example to show that the EFE contains solutions counterintuitive to our contemporary concept of "absolute" time; his model admits closed time-like curves (CTCs), violating causality in the Universe. Nevertheless, it spurred a series of studies in the following decades, such as those by \cite{silk1966,silk1970,hawking1969,hawking1973} and others. It became evident that the causality issues can be addressed, while retaining the possibility of cosmic rotation \citep{obukhov1992}.

Recent observations of the Cosmic Microwave Background (CMB) radiation made by the Planck satellite clearly indicate that the Universe is isotropic within $10^{-5}\,K$ around its mean temperature \citep{planck2018-vi}. These measurements can tightly constrain rotation through the effects caused by shear (anisotropic rotation), as it would introduce a preferred direction, manifesting as an anisotropy in the CMB, in addition to the rotational and orbital motion of the Earth and larger systems it is part of \citep{barrow1985}. However, these types of constraints still allow isotropic models with global vorticity, as proposed by \citep{obukhov1992}.

Planck measurements were tested against homogeneous models with anisotropic rotation, collectively referred to as Bianchi cosmologies \citep{saadeh2016}. Planck eventually constrained out these cosmologies \citep{planck2015-xviii}, placing a stringent upper limit on rotation to $(\omega/H)_{0} < 7.6 \times 10^{-10}$, leaving only those models permitted that ensure a shear-free or parallax-free Universe, i.e. G{\"o}del-like metrics \citep{obukhov2000}.

The notion of 'rotation' becomes conceptually challenging when we consider the Universe in its entirety. \cite{gamow1946} original proposed that the axis of rotation may be out of the scope of our telescopes, while \cite{hawking1969} only considers rotation to be the vorticity of objects nearby\footnote{Nearby compared to the Hubble radius, but far away with respect to the galactic scale.} to the observer. As \cite{hawking1969} assesses, there is no need to designate a global axis of rotation if said rotation is spatially homogenous; if all observers see a "local" vorticity, then the Universe effectively rotates. Others, e.g. \cite{pathria1972} argued that the Universe might be the interior of a black hole with its Schwarzschild radius equal to the Hubble radius. If we consider it to be a Kerr black hole, spinning close to its maximal possible angular velocity -- just like all observed black holes \citep{daly2019} --, then we have a rotating (black hole) Universe.

Despite the elusive nature of the definition with diverse interpretations in the literature, in this article, we consider an infinite and homogenous Universe, where the vorticity of the galaxies around any observer is similar and non-zero. We also consider the rotation to be shear-free and parallax-free, attributes that are still permitted by the most recent CMB measurements.

\section{Simulation} \label{sec:simulation}

We examine the effects of a shear-free, rigid-body rotation of the Universe on its expansion rates using numerical N-body simulations. This rotation is modeled around a central observer, effectively representing a global vorticity field at cosmological scales. However, simulating such a rotating Universe presents significant technical challenges.

State-of-the-art N-body simulations typically utilize a toroidal topology, where the Universe is represented as a periodic cube \citep{efstathiou1985}. Such simulations are only capable of simulating a finite portion of the Universe through the introduction of said periodic boundary conditions and approximating force calculation schemes \citep{ewald1921,hernquist1991}, primarily for numerical convenience.

In a toroidal Universe, a global rotation would appear as a constant velocity field, which would not possess any acceleration, as we would expect from a rotating object. To overcome this limitation, we use the \texttt{StePS}\footnote{\url{https://github.com/eltevo/StePS}} (cosmological) N-body code from \cite{racz2019} that is capable of simulating the entire infinite Universe. The approach of \texttt{StePS} is based on the stereographic projection of a sphere onto a plane. Generalizing this projection to a higher dimension, we are able to compactify the entire 3D Universe onto a 4D hypersphere, where the Universe is represented as a finite volume with isotropic boundaries, which coincides with the physical expectations \citep{racz2018}.

\subsection{Initial conditions} \label{subsec:initial-conditions}

We generate a spherical, homogenous distribution of particles in a 'glass' configuration, which is a perfect lattice of particles with no gravitational force acting on them. Although cosmological simulations typically start with additional perturbations applied to the glass to seed large-scale structure (LSS) formation, in this article we choose to study an unperturbed glass only to isolate the effects of rotation from the LSS formation. Cosmological perturbations in a rotating simulation are left for future work. Characteristics of the simulations are summarized in Table \ref{tab:sim}.

Taking the glass described above, we apply the Hubble flow -- an outward radial velocity $v_{\text{Hubble}} = H_{0} \cdot r$ to the particles -- and various degrees of global rotation. Latter is applied by giving each particle an angular velocity $\omega_{init}$ around the $z$-axis in the coordinate system of the simulation.

Due to expansion and momentum conservation, rotation decays over time by a factor of $a^{-2}$, meaning we choose the initial angular velocity of the particles to be

\begin{equation*}
    \omega_{\text{init}}
    =
    \omega_{0} \cdot \left( 1 + z_{init} \right)^{2}\,,
\end{equation*}

where $\omega_{0}$ is the angular velocity at $z = 0$, i.e. the present time and $z_{init}$ is the initial redshift.

We require that rotational velocities are less than the speed of light within the horizon. This is approximately equivalent of having no closed time-like curves within our Newtonian simulation. The maximal value of $\omega_{0}$ thus can be estimated from the assumption that for $\Omega$, the angular velocity of the Universe at a given time, $\Omega \ll H_{0} \sqrt{a}$ always holds. The corresponding upper limit is $\omega_{0} \approx 10^{-3}$ Gyr$^{-1}$ for the present time.

Besides the non-rotating case of ${\omega_{\text{init}} = \omega_{0} = 0}$, we run simulations with $5$ different degrees of rotation, namely
\begin{equation*}
    \omega_{0}
    =
    \left\{
        10^{-5}, 5\times10^{-5}, 10^{-4}, 5\times10^{-4}, 10^{-3}
    \right\},
\end{equation*}
all in units of Gyr$^{-1}$. Although, in this study we only discuss the difference between the "stationary" simulation and the one with the maximum possible angular velocity $\omega_{0} = 10^{-3}$ Gyr$^{-1}$. We expect the vorticity to have a stronger effect at the earlier stages of the Universe, while decaying to negligible levels at later times.

\begin{table}[h]
    \begin{tabular*}{\columnwidth}{@{\extracolsep\fill}llll@{\extracolsep\fill}}
        \toprule
        D$_{\text{sim}}$ [Mpc] & N$_{\text{part}}$ & M$_{\text{part}}$ [M$_{\odot}$] & z$_{init}$ \\
        \midrule
        $1000$ & $15 \times 2^{16}$ & $\left[ 5\times10^{11}, 10^{16} \right]$ & $63$ \\
        \botrule
    \end{tabular*}
    \caption{The relevant simulation parameters used throughout all simulations, such as the diameter of the simulation volume in the Euclidean space $D_{\text{sim}}$, the number of particles $N_{\text{part}}$, the variable mass of the particles given as an approximate interval $M_{\text{part}}$, and the initial redshift $z_{init}$.
    } \label{tab:sim}
\end{table}

\subsection{Cosmology} \label{subsec:cosmology}

To get a more robust understanding of the effects of rotation, we run our simulations in both an Einstein--de Sitter (EdS) and a $\Lambda$CDM cosmological setting.

We use the EdS model -- a special case of the Friedmann--Lema\^{i}tre--Robertson--Walker (FLRW) Universe with a vanishing curvature and cosmological constant -- as a form of baseline model that helps us to verify our results obtained from the $\Lambda$CDM simulations. Comparing the two cosmologies, we expect to see a stronger anisotropy between the parallel and perpendicular expansion rates in the $\Lambda$CDM model compared to the EdS model.

The relevant cosmological parameters we used for all simulations can be found in Table \ref{tab:cosmo}.

\begin{table}[h]
    \begin{tabular*}{\columnwidth}{@{\extracolsep\fill}c|llllll@{\extracolsep\fill}}
        \toprule
        Cosmology & $\Omega_{m}$ & $\Omega_{r}$ & $\Omega_{k}$ & $\Omega_{\Lambda}$ & $H_{0}$ \\
        \midrule
        EdS          & $1$      & $0$ & $0$ & $0$      & $67.66$ \\
        $\Lambda$CDM & $0.3089$ & $0$ & $0$ & $0.6911$ & $67.66$ \\
        \botrule
    \end{tabular*}
    \caption{The relevant main cosmological parameters, i.e. the matter density $\Omega_{m}$, radiation density $\Omega_{r}$, curvature density $\Omega_{k}$, and the cosmological constant $\Omega_{\Lambda}$, as well as the Hubble constant $H_{0}$ used throughout the simulations. While density parameters of the EdS model are exact, for the $\Lambda$CDM model and $H_{0}$ we used the currently known best values of cosmological parameters as measured by the Planck telescope from the CMB radiation \citep{planck2018-vi}.} \label{tab:cosmo}
\end{table}

\subsection{Counteracting the emerging curvature} \label{subsec:emerging-curvature}

Expansion in a rotating Universe can be described using two different scale factors, one associated with the parallel ($a_{\parallel}$) and one with the perpendicular ($a_{\bot}$) components of the velocity field in relation to the axis of rotation. This direction-dependence introduces a term in the Friedmann equation that behaves similarly to spatial curvature. However, this is not a physical curvature of spacetime but an apparent one resulting from the rotational motion.

Nevertheless, modern cosmological surveys indicate that the Universe is flat. To compensate for the emerging curvature and 'flatten' the rotating simulations, we scale the angular velocity components when setting up the initial conditions. Since this curvature originates from the kinetic energy (KE) of the rotation, our goal is to restore the total KE of the system to match the case of the same Universe without rotation, containing only the Hubble flow.

The total KE can be described for both the rotating (r) and non-rotating (nr) cases in terms of the parallel ($\parallel$) and perpendicular ($\bot$) -- with respect to the axis of rotation -- velocity components as

\begin{equation} \label{eq:total-ke}
    \mathrm{KE}
    =
    \frac{1}{2} \sum_{i} M_{i} \left( V_{\|,\,i}^{2} + V_{\bot,\,i}^{2} \right)\,,
\end{equation}

However, if we scale both the parallel and perpendicular components equally, this anisotropy would remain. Due to the nature of rigid body rotation, parallel components are not affected by the rotation.

Hence, we are looking for some scalar $s$ to adjust the perpendicular velocities in the rotating simulation, ensuring that the total kinetic energy matches that of the non-rotating case. From Eq.~\eqref{eq:total-ke}, we can arrive to the obvious solution for $s$ as

\begin{equation*}
    s
    =
    \frac{V_{\bot,\,nr}^{2}}{V_{\bot,\,r}^{2}}\,,
\end{equation*}

where $V_{\bot,\,nr}^{2}$ and $V_{\bot,\,r}^{2}$ are the mean of the squared perpendicular velocities for the non-rotating and rotating cases, respectively, before adjusting the values of the latter. During the generation of initial conditions, after applying the Hubble flow and adding angular velocity $\omega_{init}$ to the particles, the perpendicular velocities are scaled by $s$ to compensate for the apparent curvature introduced by rotation as

\begin{equation}
    V_{\bot,\,\text{r},\,i}^{\text{scaled}}
    =
    \sqrt{s} \cdot V_{\bot,\,\text{r},\,i}^{\text{naive}}\,.
\end{equation}

\section{Results} \label{sec:results}

We measure the expansion rates of a rotating Universe both parallel and perpendicular to its axis of rotation, searching for potential anisotropies in the cosmic expansion. First, we calculate the cosmological scale factor -- i.e. the average expansion (or contraction) of distances for a set of particles -- for each cosmological simulation. Additionally, we also determine the Hubble parameter in the direction parallel and perpendicular to the axis of rotation.

To compare the expansion at any given time $t$, we need a reference point at some time $t_{0}$, which we selected as the initial conditions. The scale factor is then defined as the average of the ratio of the distance of each particle at $t$, i.e $r_i(t)$ to the distance at $t_{0}$, i.e $r_i(t_{0})$, weighted by the mass of the particles $m_{i}$. Additionally, this value is normalized by the total mass $\sum_{i} m_{i}$. This gives us the value of the scale factor in a cosmological simulation for any given time $t$ as

\begin{equation} \label{eq:scale-factor-sim}
    a(t)
    =
    \frac{1}{N \sum_i m_i}
    \cdot
    \left( \sum_i m_i \cdot \frac{r_i(t)}{r_i(t_{0})} \right).
\end{equation}

\begin{figure}
    \centering
    \includegraphics{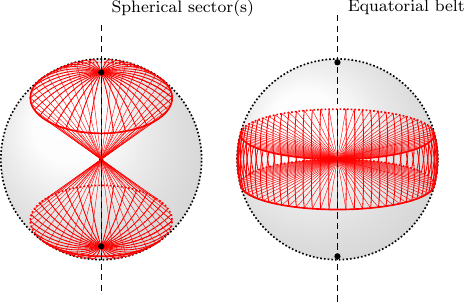}
    \caption{A schematic representation of the regions used to measure the scale factor in a rotating Universe. The left panel shows the two spherical sectors used to measure the scale factor parallel to the rotation axis, while the right panel shows the "equatorial belt" used to measure the scale factor perpendicular to the rotation axis. This is a solid of revolution that we can obtain by rotating a spherical sector perpendicular to the rotation axis. The "equatorial belt" is effectively the absolute complement of two spherical sectors inside the sphere, but with a distinct opening angle.}
    \label{fig:spherical-sectors}
\end{figure}

In regular cosmological simulations, the scale factor is always calculated with isotropy in mind, so for all particles at the same time. However, in a rotating Universe, the expansion is anisotropic, as detailed in Sec.~\ref{subsec:emerging-curvature}. This necessitates a more thoughtful approach to measuring the scale factors both in the parallel and perpendicular directions.

We calculate the scale factor parallel to the rotation axis in two joint, closed spherical sectors, both centered around the rotation axis, but in opposite directions (see left panel of Fig. \ref{fig:spherical-sectors}). Similarly, the scale factor in the perpendicular direction is measured in an "equatorial belt" (see right panel of Fig. \ref{fig:spherical-sectors}).

We select these regions in a way that the volume of the two spherical sectors combined equals to the volume of the equatorial belt, meaning

\begin{equation*}
    2 \times V_{\text{sector}} = V_{\text{belt}}.
\end{equation*}

\begin{figure}[ht]
    \centering
    \includegraphics[width=\columnwidth]{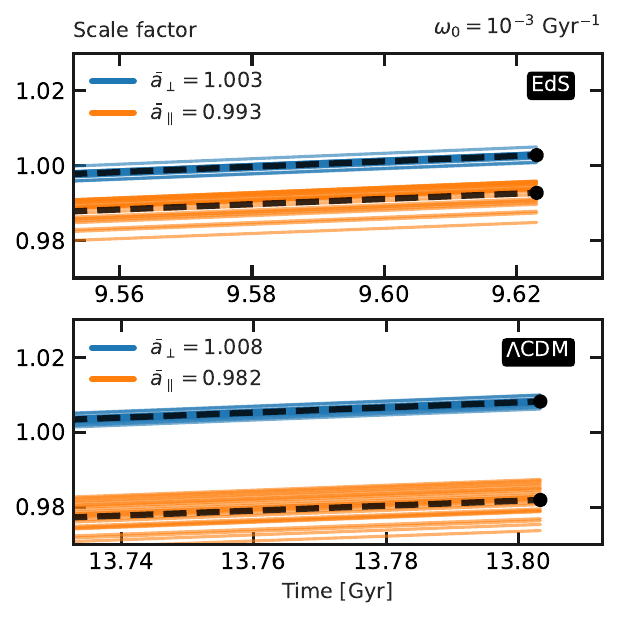}
    \caption{Comparison of scale factor evolution, in both the EdS (top panel) and $\Lambda$CDM (bottom panel) models. The figures present the results for a rotating scenario with the maximum possible angular velocity as indicated in the upper right corner. Here, only the tail of the scale factor evolution is shown, with simulations of both cosmologies ending at $z=0$, i.e. the present day. (Note the time difference between the EdS and $\Lambda$CDM universes.) The lower ensemble of curves (here, represented with solid, orange-colored lines) corresponds to the scale factors $a_{\parallel}$ calculated from polar spherical sectors with various opening angles, while the upper ensemble (solid, blue-colored lines) represents the scale factors $a_{\bot}$ in the equatorial belts. The two dashed lines indicate the mean of the two sets of measurements, $\bar{a}_{\parallel}$ and $\bar{a}_{\bot}$.}
    \label{fig:scale-factor-compare-eds-lcdm}
\end{figure}

This way, we can compare the scale factor evolution in the parallel and perpendicular directions directly. We calculate the scale factor for particles inside these regions for various opening angles of the cones and the belt as given by Eq.~\eqref{eq:scale-factor-sim}, then we aggregate them by taking the mean of the measurements. Results shown on Fig.~\ref{fig:scale-factor-compare-eds-lcdm} indicate that the scale factor in the parallel direction is always smaller than in the perpendicular direction, as expected.

\begin{figure}[ht]
    \centering
    \includegraphics[width=\columnwidth]{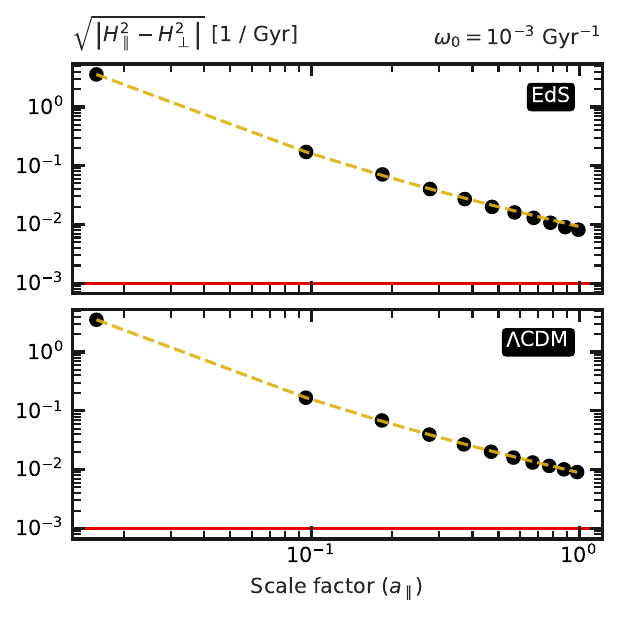}
    \caption{Comparison of the difference of the Hubble parameter between the parallel and perpendicular directions $H_{\parallel}^{2} - H_{\bot}^{2}$, in both the EdS (top panel) and $\Lambda$CDM (bottom panel) models. The figures present the square root of the results for a rotating scenario with the maximum possible angular velocity as indicated in the upper right corner. As discussed in Sec.~\ref{sec:discussion}, the square root of the difference does not converge to the expected value of $\omega_{0}$ at the present time, marked with a solid red line in the figure. Furthermore, we fit a $c_{1}/a_{\parallel}^{2} + c_{2}/a_{\parallel}$ curve to the data for both cosmological models separately, and we plot the best fit as a dashed line. These originate from the rotation and curvature terms in the rotating Newtonian Friedmann equation, respectively, where $c_{1}^{2} \equiv R$ and $c_{2}^{2} \equiv K$, with $R = \omega_{0}$ in theory.}
    \label{fig:hubble-diff}
\end{figure}

The Hubble parameter is measured in the same regions as the scale factor. The difference between the squares of the parallel and perpendicular components

\begin{equation*}
    H_{\parallel}^{2} - H_{\bot}^{2}
\end{equation*}

is then calculated and it is shown on Fig.~\ref{fig:hubble-diff}. The values for both Hubble parameters are obtained from the radial velocities of the particles in the corresponding regions using the Hubble-Lema\^{i}tre law. Furthermore, calculating the ratio of the Hubble parameters in both directions reveals an anisotropy of approximately $1~\%$.

\section{Discussion} \label{sec:discussion}

The results of our simulations show that the evolution of a rotating Universe is significantly affected by a global vorticity field. The rotation introduces two different scale factors that describe the expansion rates, which is not present in the non-rotating, true isotropic case. This leads to an expansion history, where the average mass density of the Universe $\varrho \propto a_{\parallel}^{-1} a_{\bot}^{-2}$, instead of the usual $\varrho \propto a^{-3}$, with a measured anisotropy of around $1~\%$, which is a significant deviation from the concordance model in the era of precision cosmology.

However, the exact description of this system is yet to be fully understood. We modified the Newtonian Friedmann equation by describing a homogeneous dust sphere in a rotating reference frame with some global, homogeneous vorticity. This results in a difference between the parallel and perpendicular components of the expansion rates that can be expressed as

\begin{equation*}
    H_{\bot}^{2} = H_{\parallel}^{2} + \Omega^{2}\,.
\end{equation*}

However, this approach assumes that rotation only affects the direction perpendicular to the axis of rotation. In reality, the dependence of $\varrho$ in a rotating system implies a 'back-reaction' effect of rotation in all directions. Simulations prove the existence of this effect, as seen on Fig.~\ref{fig:hubble-diff}, where $H_{\bot}^{2} - H_{\parallel}^{2}$ does not converge to $\omega_{0}^{2}$ at present. This highlights the issue that an additional term is missing from the Friedmann equation that is needed for the full description. This term is curvature-like (c.f. Fig~\ref{fig:hubble-diff}), suggesting that our naive method of removing curvature in Sec.~\ref{subsec:emerging-curvature} is unsatisfactory.

In a follow-up study, we will characterize the missing curvature term in the Newtonian rotating Friedmann equation that is relevant to a Newtonian rotating simulation.

\backmatter

\bmhead{Acknowledgements}

This work is supported by the Hungarian Ministry of Innovation and Technology NRDI Office grant OTKA NN147550, the KDP-2021 program of the Ministry of Innovation and Technology from the source of the NRDI fund and by the the European Union project RRF-2.3.1-21-2022-00004 within the framework of the MILAB Artificial Intelligence National Laboratory.

We would like to thank Prof. István Csabai and Gergely Gábor Barnaf\"{o}ldi for the regular consultations and their valuable comments and suggestions.

Simulations were carried out on the \textit{Ampere} A100 GPU server of the Wigner Scientific Computing Laboratory (WSCLAB) at the HUN-REN Wigner Research Centre for Physics, Hungary.

\bmhead{Data availability}
This study used simulated data generated using the \texttt{StePS} open-source software, which is publicly available and referenced in the article. The specific dataset used in this study was not separately archived as it can be fully reproduced using the methods and parameters described in this article.

\bibliography{./bibliography}

\end{document}